\documentclass{ijuc}
\usepackage[pdftex]{graphics}

\begin{document}

\title{Quantum Ternary Circuit Synthesis Using Projection Operations}

\author{Sudhindu Bikash Mandal\inst{1}\email{sudhindu.mandal@gmail.com}
\and Amlan Chakrabarti$^1$ 
\and \\Susmita Sur-Kolay\inst{2}
}

\institute{A. K. Choudhury School of Information Technology, University of Calcutta, India
\and
Advanced Computing \& Microelectronics Unit, Indian Statistical Institute, India
}

\maketitle

\begin{abstract}
	Basic logic gates and their operations in ternary quantum domain are involved in the synthesis of ternary quantum circuits. Only a few works define ternary algebra for ternary quantum logic realization. In this paper, a ternary logic function is expressed in terms of projection operations including a new one. A method to realize new multi-qutrit ternary gates in terms of generalized ternary gates and projection operations is also presented. We also introduced ten simplification rules for reducing ancilla qutrits and gate levels. Our method yields lower gate cost and fewer gate levels and ancilla qutrits than that obtained by earlier methods for the ternary benchmark circuits. The $n$ qutrit ternary sum function is synthesized without any ancilla qutrit by our proposed methodology.
\end{abstract}


\section{Introduction}

Multi-valued quantum computing is gaining importance in the field of quantum information theory and quantum cryptography as it can represent an \emph{n}-dimensional quantum system, defined by the basis states $\vert 0>$, $\vert 1>$, ..., $|n-1>$. The unit of information is called a qudit. A qudit exists as a linear superposition of states, and is characterized by a wave function $\psi$ \cite{N02,Ek02}.

Multi-valued quantum algebra comprises the rules for a set of basic logic operations that can be performed on qudits. While in \cite{MS00} the structure of a multi-valued logic gate is proposed which can be experimentally feasible with a linear ion trap scheme for quantum computing, this approach produces large dimensional circuits. A universal architecture for multi-valued reversible logic is given in \cite{P00}, but quantum realization of the circuits thus obtained is not apparent. The universality of $n$-qudit gates is presented in \cite{B02}, but no algorithms for synthesis were given. Al-Rabedi proposed in \cite{P01} the minimization technique for multi-valued quantum galois field sum of product. 

In this paper we concentrate on the quantum ternary functions. A ternary quantum system exists in linear superposition of three basis states, labeled $\vert 0>$, $\vert 1>$ and $\vert 2>$. All the operations on a qutrit are developed in a 3-dimensional Hilbert space under the field GF(3) \cite{K04}.  A two qutrit vector can be represented as $\vert\psi>$ = $\sum^8_{i=0}|C_i|^{2}$ = 1. In general, a $n$ qutrit state can be represented as a superposition of $3^n$ basis states. A quantum register of size $n$ qutrits can hold $3^n$ values simultaneously, whereas the $n$-qubit register in binary quantum domain can hold $2^n$ values. For the same size of memory, a Quantum Fourier Transform (QFT) with ternary qutrits improves the approximation and increases the state space by a factor of $(3/2)^n$ \cite{ZR07}. This is very useful in cryptography. 

The realization of ternary operations requires a set of universal gates with which a quantum circuit for a given ternary quantum function can be built. One set of universal gates  \cite{Y05} comprise the Ternary Swap, the Ternary NOT and the Ternary Toffoli gate. With these universal gates, any arbitrary quantum circuit can be realized without ancilla bits, but this approach is not optimal in terms of quantum cost \cite{Y05}. Further, the synthesis method in \cite{G07} uses quantum multiplexers and the method of iterative deepening depth first search is employed to minimize the gate cost. Another synthesis technique was proposed in \cite{Kh09} for garbage free GF(3) based reversible or quantum logic circuit from its truth values, using only Muthukrishnan-Stroud (M-S) [3] and shift gates \cite{P02}. But this technique did not provide any simplification rule to reduce the gate count. In paper \cite{P04} the ternary quantum logic was expressed in terms of Ternary Galois Field Sum of Product, and 16 Ternary Galois Field expansions were also proposed. But this paper did not provide any gate level implementation of ternary benchmark circuit.

The paper here is the extended version of \cite{MCS11}, where our motivation is to synthesis a given quantum ternary benchmark \cite{P04} function specified as ternary minterms, using projection operations. Further we realize the gate level implementation of these benchmark \cite{P04} circuits using Generalized Ternary Gates \cite{ZR07} and newly defined {\it permutative} ternary quantum C$^2$NOT gates. A few simplification rules are enumerated for reducing the number of ancilla qutrits and the levels of gates  in the resulting quantum circuit.

The preliminary concepts of ternary algebra with a new projection operation appear in Section 2.  In Section 3, the characteristics of basic ternary logic gates and the Generalized Ternary Gate (GTG) \cite{P02} along with the introduction of a few new gates are discussed. The proposed synthesis methodology along with a set of simplification rules is presented in Section 4. Synthesis of quantum ternary benchmark circuits by this method, and the respective quantum gate cost with ancilla qutrit count are given in Section 5. Concluding remarks appear in Section 6.

\section{Ternary Algebra}
We first define the basic operations over the set \{0, 1, 2\}. 
\subsection{Ternary AND, OR, NOT}

The operations of AND, OR, and NOT \cite{H68} for ternary variables are:\\
$AND(a,b)=\left\{\begin{array}{ll}\mbox{$a$} & \mbox{if $a \leq b$};\\
\mbox{$b$} &  \mbox{otherwise}.\\
\end{array}\right. $
$OR(a,b)=\left\{\begin{array}{ll}\mbox{$a$} & \mbox{if $a \geq b$};\\
\mbox{$b$} &  \mbox{otherwise}.\\
\end{array}\right. $\\
$NOT(a)=a+1$.\\
Here $'+'$ denotes the modulo 3 addition. It is easy to see that the ternary $AND$ operation is distributive over the $OR$ operation and vice-versa \cite{H68}.

\subsection{Ternary Projection Operations $L_i$ and $J_i$}
Our main goal is to define a given ternary function in terms of the minterms corresponding to its $1$ or $2$ values. The $3^n$ minterms for a function $f$ with $n$ ternary variables are denoted by $m_0$ to $m_{3^n-1}$. In the sum of products form of minterms, the minterm $m_0$ = $\prod^n_{i}NOT(x_i)$ where $x$ = \{$0$, $1$, $2$\} and $\prod$ denotes the ternary $AND$; its value is 1, 2, or 0 depending on whether for all \emph{i} $x_i$ = 0, 1 or 2 respectively. 

As the outcomes of a ternary function may be $0$, $1$ or $2$, we cannot express a ternary function $f$ which results in either $1$ or $2$ for a given set of inputs, by using the above minterms $m_i$. So, we present six projection operations, grouped into two types $L_i$ and $J_i$, where $i$ = \{$0$, $1$, $2$\}. While the $J_i$ type operation was defined earlier \cite{H68} as 


$J_i(a)=\left\{\begin{array}{ll}\mbox{$2$} & \mbox{if $a = i$};\\                                                      \mbox{$0$} &  \mbox{otherwise}.\\
\end{array}\right. $\\

the $L_i$ types are newly defined here as 

$L_i(a)=\left\{\begin{array}{ll}\mbox{$1$} & \mbox{if $a = i$};\\                                                      \mbox{$0$} &  \mbox{otherwise}.\\
\end{array}\right. $\\

Their truth tables appear in Table 1. From Table 1, we can verify that the output of any $L_i$ operation is either $0$ or $1$, and that of $J_i$ operation is either $0$ or $2$. Two other derived operations $L^\prime_{i}$, $J^\prime_{i}$ as shown in the Table 2. Next, we express a ternary function as a sum of products minterm form by using the $L_i$ and $J_i$ operations.  The $L_i$ and $J_i$ operations are commutative, associative and distributive over ternary AND and OR logic. We can verify these laws given below, from Table 1. Here $i$, $j$, $k$ = \{0, 1, 2\}\\
1. Commutativity\\
\begin{tabular}{l l}
(i) & $L_i(a)+L_j(b)$ = $L_j(b)+L_i(a)$ \\ &  $J_i(a)+J_j(b)$ = $J_j(b)+ J_i(a)$\\
(ii) & $L_i(a).L_j(b)$ = $L_j(b). L_i(a)$ \\ &  $J_i(a).J_j(b)$ = $J_j(b).J_i(a)$\\
\end{tabular}\\
2. Associativity\\
\begin{tabular}{l l}
(i) & [$L_i(a)+L_j(b)$]+ $L_k(c)$ = $L_i(a)$+ [$L_j(b) + L_k(c)$] \\ &  [$J_i(a)+J_j(b)$]+ $J_k(c)$ = $J_i(a)$+ [$J_j(b) + J_k(c)$]\\
(ii) & [$L_i(a).L_j(b)$].$L_k(c)$ = $L_i(a)$. [$L_j(b).L_k(c)$] \\ &  [$J_i(a).J_j(b)$].$J_k(c)$ = $J_i(a)$. [$J_j(b).J_k(c)$]\\
\end{tabular}\\
3. Distributivity\\
(i)\hspace*{3 mm}$L_i(a)$.[$L_i(b)+L_k(c)$]=$L_i(a).L_j(b) + L_i(a).L_k(c)$ \\ \hspace*{3.5 mm}  $J_i(a)$.[$J_i(b)+J_k(c)$]=$J_i(a).J_j(b) + J_i(a).J_k(c)$\\
(ii)  $L_i(a)$+ [$L_j(b)+L_k(c)$] = [$L_i(a)+ L_j(b)$].[$L_i(a)+ L_k(c)$] \\ \hspace*{3 mm} $J_i(a)$+ [$J_j(b)+J_k(c)$] = [$J_i(a)+ J_j(b)$].[$J_i(a)+ J_k(c)$]\\
\begin{table}
\small
\begin{center}
\begin{tabular}{|c|c|c|c||c|c|c|}
\hline
$a$ & $L_0(a)$&$L_1(a)$&$L_2(a)$&$J_0(a)$&$J_1(a)$&$J_2(a)$\\
\hline
0 & 1&0&0&2&0&0\\
1 & 0&1&0&0&2&0\\
2 & 0&0&1&0&0&2\\
\hline
\end{tabular}
\end{center}
\caption{Truth tables of ternary projection operations $L_i$ and $J_i$ }
\end{table}

\begin{table}[h]

\begin{center}
\begin{tabular}{|c|c|c|c||c|c|c|}
\hline
$a$ & $L^\prime_{0}(a)$&$L^\prime_{1}(a)$&$L^\prime_{2}(a)$&$J^\prime_{0}(a)$&$J^\prime_{1}(a)$&$J^\prime_{2}(a)$\\
\hline
0 & 0&1&1&0&2&2\\
1 & 1&0&1&2&0&2\\
2 & 1&1&0&2&2&0\\
\hline
\end{tabular}
\end{center}
\caption{Truth tables of ternary operations $L^\prime_{i}$ and $J^\prime_{i}$}
\end{table}

\normalsize
\section{Ternary Logic Gates}
\subsection{Ternary Feynman, Ternary Toffoli, MAX and MIN gates}
The two qutrit ternary Feynman gate \cite{Kh02}, shown in Figure 1, is defined as \\
Feynman($A$, $B$) = $A$ + $B$ \\
where '+' operator is addition over GF(3). 

The ternary 3-qutrit Toffoli gate \cite{Kh02} in the universal ternary quantum get set, is shown in Figure 2 and is defined as \\
$Toffoli($A$, $B$, $C$)=\left\{\begin{array}{ll}\mbox{$Z$ operation on $C$} & \mbox{\emph {if A = B = 2}};\\
\mbox{$C$} &  \mbox{\em otherwise};\\
\end{array}\right.$ \\
where $A$, $B$ are controlling input and $Z$ =\{$+1$, $+2$, $01$, $02$, $12$\}. 

In our synthesis method, ternary SWAP and NOT gates are not used explicitly. Instead, we use the ternary MAX and MIN gates \cite{G07} respectively to replace the OR and the AND in ternary quantum logic. These two gates are defined as\\
$MAX(A_1,A_2,..,A_n, $B$)=\left\{\begin{array}{ll}\mbox{$A_i$} & \mbox{\em if $A_i$ $\geq$ $A_j$, $i \neq j$ and  $A_i$ $\geq$ $B$};\\
\mbox{$B$} &  \mbox{\em if $\forall i, B \geq A_i$};\\
\end{array}\right.$ \\
\\
$MIN(A_1,A_2,...A_n, $B$)=\left\{\begin{array}{ll}\mbox{$A_i$} & \mbox{\em if $A_i$ $\leq$ $A_j$, $i \neq j$ and $A_i \leq B$};\\
\mbox{$B$} &  \mbox{\em if $\forall i, B \leq A_i$};\\
\end{array}\right. $\\
where $i$ = $\{1,2, \ldots n\}$ (Figure 3). 

The ternary Toffoli gate can be realized without any ancilla qutrits \cite{Kh02, Kh09b} by using  Muthukrishnan-Stroud gates (M-S gate) and . An M-S gate is defined as \\ 
{\it M-S}$(A, B)=\left\{\begin{array}{ll}\mbox{B shift by Z} & \mbox{\em if A = 2};\\
\mbox{B} &  \mbox{\em otherwise};\\
\end{array}\right. $\\ 
where $Z$ = \{$+1$, $+2$, $01$, $02$, $12$\} \cite{H68} (Figure 4(a)).

\begin{figure}
\begin{center}
\scalebox{0.6}{\includegraphics{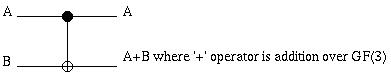}}
\end{center}
\caption{2 qutrit ternary Feynman Gate}
\label{fig-eg}
\end{figure}
\begin{figure}
\begin{center}
\scalebox{0.5}{\includegraphics{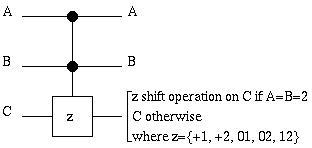}}
\end{center}
\caption{3 qutrit ternary Toffoli Gate}
\label{fig-eg}
\end{figure}

\begin{figure}
\begin{center}
\scalebox{0.5}{\includegraphics{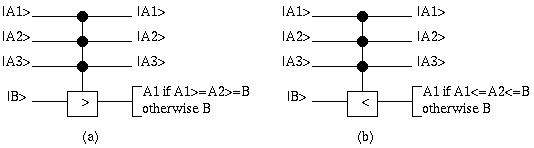}}
\end{center}
\caption{Ternary Multi-qutrit (a) $MAX$ gate, and (b) $MIN$ gate}
\label{fig-eg}
\end{figure}

\begin{figure}
\begin{center}
\scalebox{0.5}{\includegraphics{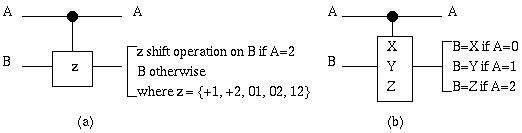}}
\end{center}
\caption{(a) A Muthukrishnan-Stroud (M-S) gate, and (b) a Generalized Ternary Gate (GTG)}
\label{fig-eg}
\end{figure}

\begin{table}
\small
\caption{Ternary Shift operations used in a GTG}
\begin{center}
\begin{tabular}{|l|l|}
\hline
Shift gate & Its operation\\
\hline
Buffer & $x$ = $x$\\
Single Shift & $x$ = $x+1$\\
Dual Shift & $x$ = $x+2$\\
Self Shift & $x$ = $2x$\\
Self Single Shift & $x$ = $2x+1$\\
Self Dual Shift & $x$ = $2x+2$\\
\hline
\end{tabular}
\end{center}
\end{table}

\normalsize
\subsection{Generalized Ternary Gate}
In order to realize a Generalized ternary gate (GTG) defined next, the Shift gates \cite{P02} given in Table 3 are required. Here addition and multiplication over $GF(3)$ are denoted by '$+$' and '$.$' respectively.

 A Generalized Ternary Gate (GTG) is a 2-qutrit gate, as shown in Figure 4(b). The controlling input of a GTG can be used to select the 1-qutrit shift operation on the target input. The GTG is defined as\\
$GTG(A,B)=\left\{\begin{array}{lll}\mbox{B shift X} & \mbox{\em if A=0};\\
\mbox{B shift Y} &  \mbox{\em if A=1};\\
\mbox{B shift Z} &  \mbox{\em if A=2} \end{array}\right. $\\

where $X$, $Y$, and $Z$ are any three distinct shift operations in Table 3.

\subsubsection{Implementation of $L_i$ and $J_i$ operations using GTG}
We can implement the $L_i$ and $J_i$ projection operations defined in Section 2.2, by using GTG as shown in Figures 5 (a) and (b) respectively.
For the $L_i$ type operation, we set $b = 1$.\\
 $L_i(a)=GTG(a, b)=\left\{\begin{array}{lll}\mbox{b} & \mbox{\em if a=i};\\
\mbox{b+2} &  \mbox{\em if a=i+1};\\
\mbox{2b+1} &  \mbox{\em if a=i+2} \end{array}\right. $\\

For the $J_i$ type operation, we set $b = 2$.\\
 $J_i(a)=GTG(a, b)=\left\{\begin{array}{lll}\mbox{b} & \mbox{\em if a=i};\\
                                                         \mbox{b+1} &  \mbox{\em if a=i+1};\\
                                                         \mbox{2b+2} &  \mbox{\em if a=i+2} \end{array}\right. $\\
\begin{figure}
\begin{center}
\scalebox{0.5}{\includegraphics{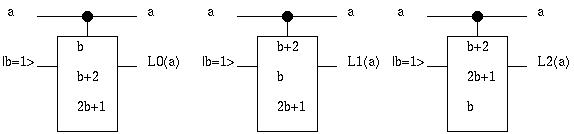}}
\end{center}
\caption{Realization of ternary (a) $L_i$ operation and (b) $J_i$ operation with GTGs}
\label{fig-eg}
\end{figure} 
 
 \begin{figure}
\begin{center}
\scalebox{0.5}{\includegraphics{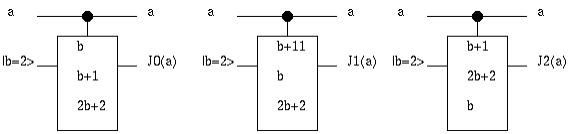}}
\end{center}
\caption{Realization of ternary $J_i$ operation wth GTG}
\label{fig-eg}
\end{figure}

\subsection{A new Ternary $C^2NOT$ gate}
Several multi-qutrit control operations are possible in ternary logic. We present a new definition of a 3-qutrit $C^2NOT$ (Figure 6(a)), which is required to realize the simplification rules for ternary minterms given in the Section 4.2 below, as\\
$C^2NOT(A,B,C)=\left\{\begin{array}{ll}\mbox{NOT(C)} & \mbox{\em if $A\neq B$ and $A$,$B\neq 0$ };\\
\mbox{C} &  \mbox{\em otherwise};\\
\end{array}\right. $\\
where $A$ and $B$ are the control inputs, and $C$ the target input.
The realization of this $C^2NOT$ gate with M-S gates is shown in Figure 6(b).


\begin{figure}
\begin{center}
\scalebox{0.5}{\includegraphics{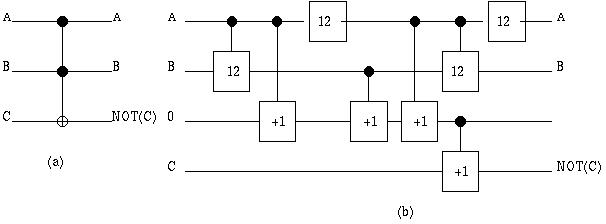}}
\end{center}
\caption{(a) The proposed ternary $C^2NOT$ gate, and (b) its realization with M-S gates}
\label{fig-eg}
\end{figure} 
\subsection{A Multi-qutrit GTG }
We also define a Multi-qutrit GTG (Figure 7) as\\
GTG($A_1, A_2,..,A_n$, B)=$\left\{\begin{array}{llll}\mbox{B shift X} & \mbox{\em if $A_1, A_2,..,A_n$=0};\\
\mbox{B shift Y} &  \mbox{\em if $A_1, A_2,..,A_n$=1};\\
\mbox{B shift Z} &  \mbox{\em if $ A_1, A_2,..,A_n$=2};\\ 
\mbox{B} &  \mbox{\em otherwise};
\end{array}\right.$ 
				         
\begin{figure}
\begin{center}
\scalebox{0.5}{\includegraphics{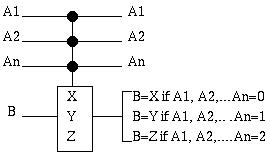}}
\end{center}
\caption{The proposed Multi-qutrit GTG}
\label{fig-eg}
\end{figure}

\section{Proposed Synthesis Methodology}
\subsection{Overview}
Consider an $m$-variable ternary quantum logic function  $f(a_1, a_2, \ldots ,a_m)$ = $\sum^n_{i=0}$(minterms for \emph{one} )$_i$ + $ \sum^p_{j=0}$(minterms for \emph{two})$_j$ \\
where $\sum$ implies logical ternary \emph{OR}, \emph{ n} and \emph{p} are respectively the number of input vectors for which $f$ is $1$, and  $2$. Thus, for $(3^m-n-p)$ input vectors, $f$ is 0 . We express the minterm for \emph{one} and \emph{two} by using the $L_i$ and $J_i$ operations respectively. From Table II, we can say \\
$\prod^m_{i=1}L_0(a_i)$ = $\left\{\begin{array}{ll}\mbox{1} & \mbox{\em if $\forall i$ $a_i$ = 0};\\  
\mbox{0} & \mbox{\em if $\exists$ i $a_i$ = 1 or 2};\end{array}\right.$ \\
 $\prod^m_{i=1}L_1(a_i)$ = $\left\{\begin{array}{ll}\mbox{1} & \mbox{\em if $\forall i$ $a_i$ = 1};\\
\mbox{0} & \mbox{\em if $\exists$ i $a_i$ = 0 or 2};\end{array}\right.$ \\
$\prod^m_{i=1}L_2(a_i)$ =$\left\{\begin{array}{ll}\mbox{1} & \mbox{\em if $\forall i$ $a_i$ = 2};\\
\mbox{0} & \mbox{\em if $\exists$ i $a_i$ = 0 or 1};\end{array}\right.$ \\ 
$\prod^m_{i,p,k=1}L_0(a_i).L_1(a_p).L_2(a_k)$ =$\left\{\begin{array}{ll}\mbox{1} & \mbox{\em if $\forall i, p, k$ $a_i$ = 0, $a_p$ = 1, $a_k$ = 2};\\
\mbox{0} & \mbox{\em if $\exists$ i, p, k  $a_i$ = 1 or 2, $a_p$ = 0 or 2, $a_k$ = 0 or 1};\end{array}\right.$\\
where $i+p+k=m$.

Hence, the minterms for which $f = 1$ are\\
1. $\prod^m_{i=1}L_0(a_i=0)$,\\
2. $\prod^m_{i=1}L_1(a_i=1)$,\\
3. $\prod^m_{i=1}L_2(a_i=2)$, and\\
4. $\prod^m_{i,p,k=1}L_0(a_i=0).L_1(a_p=1).L_2(a_k=2)$.

Similarly from Table I, the minterms for which $f = 2$ are\\  
1. $\prod^m_{i=1}J_0(a_i=0)$,\\
2. $\prod^m_{i=1}J_1(a_i=1)$,\\
3. $\prod^m_{i=1}J_2(a_i=2)$, and\\
4. $\prod^m_{i,p,k=1}J_0(a_i=0).J_1(a_p=1).J_2(a_k=2)$.

\subsection{Simplification Rules}
Next, we define seven simplification rules for our proposed method, the first four are derived from Table 1, and the next three from Table 2.\\
1. $L_i(a).0=0$, and $J_i(a).0=0$\\
2. $L_i(a).1=L_i(a)$, and $J_i(a).2=J_i(a)$\\
3. $L_i(a)+0=L_i(a)$, and $J_i(a)+0=J_i(a)$\\
4. $L_i(a)+1=1$, and $J_i(a)+2=2$\\
5. $L_i(a).L^\prime_{i}(a)=0$, and $J_i(a).J^\prime_{i}(a)=0$\\
6. $L_i(a)+L^\prime_{i}(a)=1$, and $J_i(a)+J^\prime_{i}(a)=2$\\
7. $L^\prime_{i}(a)=L_{i+1}(a)+L_{i+2}(a)$, and $J^\prime_{i}(a)=J_{i+1}(a)+J_{i+2}(a)$.

\subsubsection{Simplification Rules for reducing ancilla qutrits}
For gate level realization of the projection operations $L_i$ and $J_i$, we need an ancilla qutrit for each. Further, to synthesize an $m$-variable ternary function with $n$ minterms specified according to our proposed methodology, we have maximum of $n*m$ ancilla qutrits. However, we can reduce the number of ancilla qutrits by the following three simplification rules based on the new ternary $C^2NOT$ gate and Table 2:\\
8. $L_1(a)L_2(b)+L_2(a)L_1(b)=C^2NOT(a,b,0)$ and
$J_1(a)J_2(b)+J_2(a)J_1(b)=C^2NOT(a,b,1)$\\
9.
 $L_i(a).L_i(a)=L_i(a)$ and $J_i(a).J_i(a)=J_i(a)$\\
10. $L_i(a_1)L_i(a_2)..L_i(a_n)=L_i(a_1,a_2,..,a_n)$ and
 $J_i(a_1)J_i(a_2)..J_i(a_n)=J_i(a_1,a_2,..,a_n)$, $i=\{0,1,2\}$.\\

      The multiqutrit $L_i$ and $J_i$ operations in Rule 10 can be realized using Multiqutrit GTG as follows:\\
      For $L_i(a_1,a_2,..,a_n)$, 
$GTG(a_1,a_2,..,a_n, B)=\left\{\begin{array}{llll}\mbox{B +1} & \mbox{if $a_1,a_2,..,a_n$=i};\\
\mbox{B} &  \mbox{if $a_1,a_2,..,a_n$=i+1};\\
\mbox{2B} &  \mbox{if $a_1, a_2,..,a_n$=i+2};\\ 
\mbox{B} &  \mbox{otherwise};\end{array}\right. $\\
\\
For $J_i(a_1,a_2,..,a_n)$, 
$GTG(a_1,a_2,..,a_n, B)=\left\{\begin{array}{llll}\mbox{B +2} & \mbox{if $a_1,a_2,..,a_n$=i};\\
\mbox{B} &  \mbox{if $a_1,a_2,..,a_n$=i+1};\\
\mbox{2B} &  \mbox{if $a_1, a_2,..,a_n$=i+2};\\ 
\mbox{B} &  \mbox{otherwise};\end{array}\right. $\\

\subsection{An Example}
For illustrating our method, consider an arbitrary ternary function $g(a, b)$ with its truth table given in Table 4.

\begin{table}
\small
\caption{Truth table for an example ternary 2-qutrit function $g(a, b)$ }
\begin{center}
\begin{tabular}{|c|c|c|c|c|c|c|c|c|c|}
\hline
$a,b$ & $00$ & $01$ & $02$ & $10$ & $11$ & $12$ & $20$ & $21$ & $22$\\
\hline
$g(a,b)$ & 0 & 1 & 2 & 1 & 1 & 1 & 2 & 1 & 2\\
\hline
\end{tabular}
\end{center}
\end{table}

\normalsize
The disjunction of minterms for which $g = 1$ is \\$L_0(a)L_1(b)+L_1(a)L_0(b)+L_1(a)L_1(b)+L_1(a)L_2(b)+L_2(a)L_1(b)$ \\

and that for which $g = 2$ is\\  
$J_0(a)J_2(b)+J_2(a)J_0(b)+J_2(a)J_2(b)$.

Hence, $g(a,b)= L_0(a)L_1(b)+L_1(a)L_0(b)+L_1(a)L_1(b)+L_1(a)L_2(b)+L_2(a)L_1(b)+J_0(a)J_2(b)+J_2(a)J_0(b)+J_2(a)J_2(b)$. 
\\By simplification rules 1 and 10, we get\\
$g(a, b)=L_0(a)L_1(b)+L_1(a)L_0(b)+L_1(a,b)+C^2NOT(a,b,0)+J_0(a)J_2(b)+J_2(a)J_0(b)+J_2(a,b).$ \\The gate level implementation of the function $g(a, b)$ is shown in Figure 8. 

\begin{figure}
\begin{center}
\scalebox{0.3}{\includegraphics{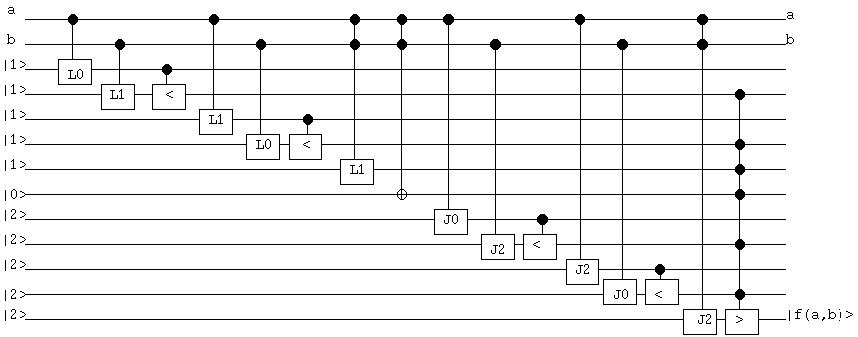}}
\end{center}
\caption{Ternary quantum gate circuit obtained for $g(a,b)$ in Table 4}
\label{fig-eg}
\end{figure} 
\section{Synthesis of Ternary Benchmark functions}
We experimented with the ternary benchmark functions provided in \cite{P04}. Their truth tables appear in Table 5. The definition and the synthesis of the ternary benchmark functions using ternary minterm are given in the following subsections.

\subsection{2-qutrit multiplication functions $mul2$ and $mul2c$}
The functions are defined as\\
$mul2(a,b) = ab mod 3$, {\em multipication output};\\
$mul2c(a,b) = int[ab/3]$, {\em carry of the multiplier}.\\
The truth table of $mul2(a,b)$ and $mul2c(a,b)$ appear in columns $3$ and $4$  respectively of Table 5. \\
Expressing in terms of projrctin operators, we get \\
$mul2(a,b)= L_1(a)L_1(b)+L_2(a)L_2(b)+J_1(a)J_2(b)+J_2(a)J_1(b)$.\\
\hspace*{17 mm}= $L_1(a,b)+L_2(a,b)+C^2NOT(a,b,1)$  (by Rules 8 and 10)\\
Similarly, $mul2c(a,b) = L_2(a)L_2(b)= L_2(a,b)$ (by Rule 10).

\subsection{2-qutrit half-adder $thadd$ }
The 2-qutrit half-adder $thadd$ \cite{MCS11} comprises the ternary half adder sum ($sumh$) and carry ($carryh$) functions, whose truth tables are given in columns 5 and 6  respectively of Table 5.  \\
We re-write $sumh$ according to our proposed methodology as\\ 
$sumh$ = $L_0(a)L_1(b)+L_1(a)L_0(b)+L_2(a)L_2(b)+J_0(a)J_2(b)+J_1(a)J_1(b)+J_2(a)J_0(b)$.\\
Further simplification of $sumh$ by using the ternary Feynman gate of Section 3.1 gives $sumh=a+b$.\\
Next, the carry function $carryh$ is written as\\
$carryh = L_1(a)L_2(b)+L_2(a)L_1(b)+L_2(a)L_2(b)= C^2NOT(a,b,0)+ L_2(a, b)$ (by Rules 8 and 10).
\hspace*{17 mm}
\begin{table}
\small
\caption{Truth table of four $2$-qutrit benchmark functions[13]}
\begin{center}
\begin{tabular}{|c|c|c|c|c|c|c|c|}
\hline
$a$ & $b$ & $mul2$ & $mul2c$ & $sumh$ & $carryh$ & $sqsum2$ & $avg2$\\
\hline
0 & 0 & 0 & 0 & 0 & 0 & 0 & 0 \\
0 & 1 & 0 & 0 & 1 & 0 & 1 & 0 \\
0 & 2 & 0 & 0 & 2 & 0 & 1 & 1 \\
1 & 0 & 0 & 0 & 1 & 0 & 1 & 0 \\
1 & 1 & 1 & 0 & 2 & 0 & 2 & 1 \\
1 & 2 & 2 & 0 & 0 & 1 & 2 & 1 \\
2 & 0 & 0 & 0 & 2 & 0 & 1 & 1 \\
2 & 1 & 2 & 0 & 0 & 1 & 2 & 1 \\
2 & 2 & 1 & 1 & 1 & 1 & 2 & 2 \\
\hline
\end{tabular}
\end{center}
\end{table}

\normalsize

\subsection{GF(3) sum of two squares $sqsum2$}
The definition of this ternary function is $sqsum2(a,b)=(a^2+b^2) mod 3$ and its truth table is shown in column 7 of Table 5.\\
$sqsum2(a,b) = L_0(a)L_1(b)+ L_0(a)L_2(b) + L_1(a)L_0(b) + L_2(a)L_0(b)+\hspace*{23 mm}J_1(a)J_1(b) +J_1(a)J_2(b)+J_2(a)J_1(b)+J_2(a)J_2(b)$\\
\hspace*{20 mm}= $L_0(a)L_0'(b)$+$L_0'(a)L_0(b)$+$J_1(a,b)+C^2NOT(a,b,2)+ J_2(a,b)$ \\
\hspace*{17 mm} (by Rules 8, 7 and 10).
\subsection{Average function $avg2$}
The function $avg2$ is the integer part of the average of two ternary input variables in $modulo 3$ and defined as $avg2(a,b)=int[ab/2]$. Its truth table is in column 8 of Table 5.\\
$avg2 = L_0(a)L_2(b) + L_1(a)L_1(b) + L_1(a)L_2(b) + L_2(b)L_0(b)+ L_2(a)L_1(b)+ J_2(a)J_2(b)$\\
\hspace*{11 mm}=$L_0(a)L_2(b)+L_1(a,b)+C^2NOT(a,b,1)+L_2(a)L_0(b)+ J_2(a,b)$
(by Rules 8 and 10).

\begin{table}
\small
\caption{Truth table of four $3$-qutrit benchmark functions [13]}
\begin{center}
\begin{tabular}{|c|c|c|c|c|c|c|c|}
\hline
$a$ & $b$ & $c$ & $mul3$ & $mul3c$ & $a2bcc$ & $avg3$ & $sqsum3$ \\
\hline
0 & 0 & 0 & 0 & 0 & 0 & 0 & 0 \\
0 & 0 & 1 & 0 & 0 & 1 & 0 & 1 \\
0 & 0 & 2 & 0 & 0 & 2 & 0 & 1 \\
0 & 1 & 0 & 0 & 0 & 0 & 0 & 1 \\
0 & 1 & 1 & 0 & 0 & 2 & 0 & 2 \\
0 & 1 & 2 & 0 & 0 & 1 & 1 & 2 \\
0 & 2 & 0 & 0 & 0 & 0 & 0 & 1 \\
0 & 2 & 1 & 0 & 0 & 0 & 1 & 2 \\
0 & 2 & 2 & 0 & 0 & 0 & 1 & 2 \\
1 & 0 & 0 & 0 & 0 & 1 & 0 & 1 \\
1 & 0 & 1 & 0 & 0 & 2 & 0 & 2 \\
1 & 0 & 2 & 0 & 0 & 0 & 1 & 2 \\
1 & 1 & 0 & 0 & 0 & 1 & 0 & 2 \\
1 & 1 & 1 & 1 & 0 & 0 & 1 & 0 \\
1 & 1 & 2 & 2 & 0 & 1 & 1 & 0 \\
1 & 2 & 0 & 0 & 0 & 1 & 1 & 2 \\
1 & 2 & 1 & 2 & 0 & 1 & 1 & 0 \\
1 & 2 & 2 & 1 & 1 & 1 & 1 & 0 \\
2 & 0 & 0 & 0 & 0 & 1 & 0 & 1 \\
2 & 0 & 1 & 0 & 0 & 2 & 1 & 2 \\
2 & 0 & 2 & 0 & 0 & 0 & 1 & 2 \\
2 & 1 & 0 & 0 & 0 & 1 & 1 & 2 \\
2 & 1 & 1 & 2 & 0 & 0 & 1 & 0 \\
2 & 1 & 2 & 1 & 1 & 2 & 1 & 0 \\
2 & 2 & 0 & 0 & 0 & 1 & 1 & 2 \\
2 & 2 & 1 & 1 & 1 & 1 & 1 & 0 \\
2 & 2 & 2 & 2 & 2 & 1 & 2 & 0 \\
\hline
\end{tabular}
\end{center}
\end{table}

\normalsize

\subsection{3-qutrit multiplication function $mul3$}
The function $mul3$ is defined as\\
$mul3(a,b,c) = abc mod 3$, {\em multipication output};\\
$mul3c(a,b)=int[abc/3]$, {\em carry of the multiplier}.\\

From the truth table of $mul3(a,b,c)$ in column 4 of Table 6, we have\\
$mul3(a,b,c) = L_1(a)L_1(b)L_1(c) + L_1(a)L_2(b)L_2(c) + L_2(a)L_1(b)L_2(c)+$\\
\hspace*{24mm}$L_2(a)L_2(b)L_1(c) + J_1(a)J_1(b)J_2(c) + J_1(a)J_2(b)J_1(c)+$\\
\hspace*{24mm}$J_2(a)J_1(b)J_1(c)+J_2(a)J_2(b)J_2(c)$\\
\hspace*{22mm}$=L_1(a)[L_1(b)L_1(c)+L_2(b)L_2(c)]+L_2(a)[L_1(b)L_2(c)+$\\
\hspace*{24mm}$L_2(b)L_1(c)]+J_1(a)[J_1(b)J_2(c)+J_2(b)J_2(c)]+J_2(a)[J_1(b)J_1(c)+$\\
\hspace*{24mm}$J_2(b)J_2(c)]$\\
By applying $L_i$ and $J_i$ operations on $mul2$ in column 3 of Table 5, we get  \\
\hspace*{24mm}=$L_1(a)L_1(mul2(b,c))+L_2(a)L_2(mul2(b,c))+J_1(a)J_2(mul2(b,c))+J_2(a)J_1(mul2(b,c))$.\\
\hspace*{24mm}$= mul2(a,mul2(b,c))$.\\
From the truth table of $mul3c(a,b,c)$ in column 5 of Table 6, we have\\
$mul3c(a,b,c) = L_1(a)L_2(b)L_2(c)+L_2(a)L_1(b)L_2(c)+L_2(a)L_2(b)L_1(c)+J_2(a)J_1(b)J_2(c)$\\
\hspace*{25 mm}$= C^2NOT(a,b,0)L_2(c)+L_2(a,b)L_1(c)+J_2(a,b,c)$ by Rules 8 and 10).

\subsection{Function $a^2bcc$}
This is an arbitrary function defined as 
$a^2bcc=(a^2+bc+c) mod 3$ with its truth table shown in column $6$ of Table $6$. Hence,\\
$a^2bcc = L_0(a)L_0(b)L_1(c) + L_0(a)L_1(b)L_2(c) + L_1(a)L_0(b)L_0(c) + L_1(a)L_1(b)L_0(c) +$\\
\hspace*{14mm}$L_1(a)L_1(b)L_2(c) + L_1(a)L_2(b)L_0(c) + L_1(a)L_2(b)L_1(c) + L_1(a)L_2(b)L_2(c)+$\\
\hspace*{13mm}
$L_2(a)L_0(b)L_0(c) + L_2(a)L_1(b)L_0(c) + L_2(a)L_2(b)L_0(c) + L_2(a)L_2(b)L_1(c)+$\\
\hspace*{13mm}$L_2(a)L_2(b)L_2(c) + J_0(a)J_0(b)J_2(c) + J_0(a)J_1(b)J_1(c) + J_1(a)J_0(b)J_1(c)+$\\
\hspace*{13mm}$J_2(a)J_0(b)J_1(c)+J_2(a)J_1(b)J_2(c)$\\
\hspace*{12 mm}$ = L_0(a,b)L_1(c) + L_0(a)L_1(b)L_2(c) + L_0'(a)L_0(b,c) + L_1(a,b)L_1'(c)+$
\hspace*{13mm}$C^2NOT(a,b,0)L_0(c) + L_1(a,c)L_2(b) + L_1(a)L_2(b,c) + L_2(a,b)L_2'(c)+ $
\hspace*{13mm}$L_2(a,b,c)+J_0(a,b)J_2(c) + J_0(a)J_1(a,c) + J_1(a,c)J_0(b) + J_2(a)J_0(b)J_1(c) + $
\hspace*{13mm}$J_2(a,c)J_1(b)$ (by Rules 7, 8, 10).

\subsection{Function $avg3$ function}
This function is the integer part of the average of three ternary input variables expressed as $modulo 3$, i.e., $avg3 =[int[a+b+c/3]] mod 3$. Its truth table is shown in column 7 of Table 6.\\
$avg3 = L_0(a)L_1(b)L_2(c) + L_0(a)L_2(b)L_1(c) + L_0(a)L_2(b)L_2(c) + L_1(a)L_0(b)L_2(c) + $
\hspace*{15mm}$L_1(a)L_1(b)L_1(c) + L_1(a)L_1(b)L_2(c) + L_1(a)L_2(b)L_0(c) + L_1(a)L_2(b)L_1(c) +$
\hspace*{15mm}$L_1(a)L_2(b)L_2(c) + L_2(a)L_0(b)L_1(c) + L_2(a)L_0(b)L_2(c) + L_2(a)L_1(b)L_0(c) + $
\hspace*{15mm}$L_2(a)L_1(b)L_1(c) + L_2(a)L_1(b)L_2(c) + L_2(a)L_2(b)L_0(c) + L_2(a)L_2(b)L_1(c)+ $
\hspace*{15mm}$J_2(a)J_2(b)J_2(c).$\\
\hspace*{12 mm}= $L_{0}(a)C^{2}NOT(b,c,0)$+$L_{0}(a)L_{2}(b,c)$+$L_{0}(b)C^{2}NOT(a,c,0)$+ $L_{1}(a,b,c)$+
\hspace*{15mm}$L_1(a)C^2NOT(b,c,0)$+$L_0(c)C^2NOT(a,b,0)$+$L_2(c)C^2NOT(a,b,o)$+$L_0(b)$
\hspace*{15mm}$L_2(a,c)$+$L_2(a)L_1(b,c)$+$L_2(a,b)L_0(c)$+$L_2(a,b)L_1(c)$+$J_2(a,b,c)$ (by Rules 8 and 10)\\
\hspace*{12 mm}= $L_2'(a)C^2NOT(b,c,0)$+$L_0(a)L_2(b,c)$+$L_0(b)C^2NOT(a,c,0)$+$L_1(a,b,c)$+
\hspace*{14mm}$L_1'(c)C^2NOT(a,b,0)$+$L_0(b)L_2(a,c)$+$L_2(a)L_1(b,c)$+$L_2(a,b)L_2'(c)$+$J_2(a,b,c)$ (by Rule 7).

\subsection{GF(3) sum of three squares $sqsum3$}
The definition is $sqsum3(a,b,c)=(a^2+b^2+c^2) mod 3$ with the truth table in column 8 of Table 6. \\
$sqsum3(a,b)$ = $L_0(a)L_0(b)L_1(c)$+$L_0(a)L_0(b)L_2(c)$+$L_0(a)L_1(b)L_0(c)$+$L_0(a)L_2(b)L_0(c)$+
\hspace*{21 mm}$L_1(a)L_0(b)L_0(c)$\hspace*{.6mm}+\hspace*{.6mm}$L_2(a)L_0(b)L_0(c)$\hspace*{.6mm}+\hspace*{.6mm}$J_0(a)J_1(b)J_1(c)$\hspace*{.6mm}+\hspace*{.6mm}$J_0(a)J_1(b)J_2(c)$+
\hspace*{21mm}$J_0(a)J_2(b)J_1(c)$\hspace*{1mm}+\hspace*{1mm}$J_0(a)J_2(b)J_2(c)$\hspace*{1mm}+\hspace*{1mm}$J_1(a)J_0(b)J_1(c)$\hspace*{1mm}+\hspace*{1mm}$J_1(a)J_0(b)J_2(c)$+
\hspace*{21mm}$J_1(a)J_1(b)J_0(c)$\hspace*{1mm}+\hspace*{1mm}$J_1(a)J_2(b)J_0(c)$\hspace*{1mm}+\hspace*{1mm}$J_2(a)J_0(b)J_1(c)$\hspace*{1mm}+\hspace*{1mm}$J_2(a)J_0(b)J_2(c)$+
\hspace*{21mm}$J_2(a)J_1(b)J_0(c)$+$J_2(a)J_2(b)J_0(c)$ \\
\hspace*{19 mm}= $L_0(a,b)L_1(c)$\hspace*{1.7mm}+\hspace*{1.7mm}$L_0(a,b)L_2(c)$\hspace*{1.7mm}+\hspace*{1.7mm}$L_0(a,c)L_1(b)$\hspace*{1.7mm}+\hspace*{1.7mm}$L_0(a,c)L_2(b)$+
\hspace*{21mm}$L_1(a)L_0(b,c)$+$L_2(a)L_0(b,c)$+$J_0(a)J_1(b,c)$+$J_0(a)C^2NOT(b,c,1)$+
\hspace*{21mm}$J_0(a)J_2(b,c)$\hspace*{.5mm}+\hspace*{.5mm}$J_1(a,c)J_0(b)$\hspace*{.5mm}+\hspace*{.5mm}$J_0(b)C^2NOT(a,c,1)$\hspace*{.5mm}+\hspace*{.5mm}$J_1(a,b)J_0(c)$+
\hspace*{21mm}$C^2NOT(a,b,1)J_0(c)$+$J_2(a,c)J_0(b)$+$J_2(a,b)J_0(c)$ (by Rules 8 and 10)\\
\hspace*{19 mm}= $L_0(a,b)L_0'(c)$+$L_0(a,c)L_0'(b)$+$L_0'(a)L_0(b,c)$+$J_0(a)J_1(b,c)$+$J_0(a)$
\hspace*{21mm}$C^2NOT(b,c,1)$+$J_0(a)J_2(b,c)$+$J_1(a,c)J_0(b)$+$J_0(b)C^2NOT(a,c,1)$
\hspace*{21mm}+$J_1(a,b)J_0(c)$+$C^2NOT(a,b,1)J_0(c)$+$J_2(a,c)J_0(b)$+$J_2(a,b)J_0(c)$ (by Rule 9)

\subsection{GF(3) product $prod4$}
The function $prod4$ is defined as $prod4(a,b,c,d)=(abcd) mod 3$.\\
$prod4(a,b,c,d)=L_1(a)L_1(b)L_1(c)L_1(d) + L_1(a)L_1(b)L_2(c)L_2(d) + L_1(a)L_2(b)L_1(c)L_2(d)+ $
\hspace*{28mm}$L_1(a)L_2(b)L_2(c)L_1(d) + L_2(a)L_1(b)L_1(c)L_2(d) + L_2(a)L_1(b)L_2(c)L_1(d) + $
\hspace*{28mm}$L_2(a)L_2(b)L_1(c)L_1(d) + L_2(a)L_2(b)L_2(c)L_2(d) + J_1(a)J_1(b)J_1(c)J_2(d) +$
\hspace*{28mm}$J_1(a)J_1(b)J_1(c)J_1(d) + J_1(a)J_2(b)J_2(c)J_2(d) + J_2(a)J_1(b)J_1(c)J_1(d) +$
\hspace*{28mm}$J_2(a)J_1(b)J_2(c)J_2(d) + J_2(a)J_2(b)J_1(c)J_2(d) + J_2(a)J_2(b)J_2(c)J_1(d) +$
\hspace*{28mm}$J_1(a)J_1(b)J_2(c)J_1(d)$\\
\hspace*{26mm}$=mul2(mul2(a,b),mul2(c,d))$ (by Rules 8, 10 and projection operation on $mul2$ in column 3 of Table 5).

\subsection{4-qutrit GF(3) addition function $sum4$}
This function is $sum4(a,b,c,d)=(a+b+c+d) mod 3$. Along the lines of the result for ternary full adder in \cite{MCS11}, we have\\  
by applying the simplification rules 8, 10 and projection operation on $sumh$ in column 5 of Table 5,  the minterms of $sum4$ as\\
$sum4(a,b,c,d)= L_0(a+b)L_1(c+d) + L_1(a+b)L_0(c+d) + L_2(a+b)L_2(c+d)+$
$J_0(a+b)J_2(c+d) + J_1(a+b)J_1(c+d) + J_2(a+b)J_0(c+d)$\\
\hspace*{25mm}$=sumh(sumh(a,b),sumh(c,d))$.

\subsection{Synthesis Results}
We synthesized the ternary benchmark circuit provided in \cite{P04} using our proposed synthesis methodology. The simplified forms of these functions by our method appear in Section 5.1 to 5.10 and the costs are summarized in Table 7. While the second column of the table indicates the maximum number of ancilla qutrits that may be required to synthesize the benchmark circuits, the third column shows the number of ancilla qutrit required after using our simplification rules in Section 4.2.
 
The fourth column shows the M-S gate count cost. The quantum gate cost of the ternary quantum benchmark circuits are evaluated in terms of the number of M-S gates used. The number of M-S gates required to realize a Feynman gate, a Toffolli and a GTG gate are 4, 5 and 5 respectively \cite{Kh09, H08}. To realize the new ternary $C^2NOT$ gate shown in Figure 7b, we need 8 M-S gates. 

For the ternary $sumn$ function, the number of ancilla qutrits is zero and the quantum gate cost is $(n-1)*4$. In the case of ternary $prodn$ function, the maximum number of ancilla qutrits that may be required is $2^{n}*n$. But by applying our simplification rules, we need only $(n-1)*3$ ancilla qutrits for the realization of $prodn$ function, with a quantum gate cost of $(n-1)*18$ for this function. 

Our proposed simplification rules reduce more than 50\%  of the ancilla qutrits for the functions $mul2$, $mul3$, $avgn$ and $sqsumn$. The comaprison of the M-S gate count cost with that in \cite{KP05} for these benchmark circuits are given in Table $7$. This establishes that although our design has a small increase in cost for the $sqsumn$ and $avgn$ circuits, it is 20\% less for all the other benchmark circuits.
 \begin{table}[h]
\small
\caption{Quantum Gate cost of Ternary Benchmark \cite{P04} Circuit}
\begin{center}
\begin{tabular}{|c|p{.75in}|p{.75in}|c|c|c|}
\hline
& & &\multicolumn{2}{|c|}{Gate Cost}\\
\hline
Circuit Name & Maximum Ancilla qutrit & Reduced Ancilla qutrit & our & \cite{KP05} \\
\hline
$sum2$ & 12 & 0 & 4 & 5\\
\hline
$sum3$ & 54 & 0 & 8 & 10\\
\hline
$sum4$ & 216 & 0 & 12 & 15\\
\hline
$sum5$ & 810 & 0 & 16 & -\\
\hline
$sum6$ & 2916 & 0 & 20 & -\\
\hline
$sum7$ & 10206 & 0 & 24 & -\\
\hline
$prod2$ & 8 & 3 & 18 & 20\\
\hline
$prod3$ & 24 & 6 & 36 & 65\\
\hline
$prod4$ & 64 & 9 & 54 & -\\
\hline
$prod5$ & 160 & 12 & 72 & -\\
\hline
$prod6$ & 384 & 15 & 90 & -\\
\hline
$prod7$ & 896 & 18 & 108 & -\\
\hline
$mul2$ & 10 & 4 & 23 & 25\\
\hline
$mul3$ & 36 & 11 & 64 & -\\
\hline
$thadd$ & 18 & 2 & 21 & 20\\
\hline
$tfadd$ & 63 & 4 & 42 & 55\\
\hline
$avg2$ & 12 & 7 & 38 & 15\\
\hline
$avg3$ & 51 & 16 & 89 & 40\\
\hline
$sqsum2$ & 16 & 7 & 38 & 10\\
\hline
$squsum3$ & 54 & 24 & 130 & 15\\
\hline
\end{tabular}
\end{center}
\end{table}

\normalsize

\section{Conclusion}
In this paper, we have proposed a methodology for logic synthesis of ternary quantum circuits. We have defined a minterm based approach of expressing  a ternary logic function by using $L_i$ and $J_i$ operations. We have also stated the simplification rules for the proposed method. By applying these simplification rules, we can realize the ancilla free $sumn$ function. Among the ternary benchmark circuits \cite{P04}, the quantum gate cost for the group of addition and multiplication functions $sumn$ and $prodn$, by our method is smaller compared to that by the earlier method in \cite{KP05}. 

\end{document}